\documentclass[preprint2]{aastex}

\usepackage{color}
\usepackage{multicol}
\usepackage{subfigure}
\usepackage{float}
\usepackage{hyperref}

\newcommand{\rhessi}[0]{{\it RHESSI}}
\newcommand{\goes}[0]{{\it GOES}}
\newcommand{\nustar}[0]{{\it NuSTAR}}
\newcommand{\foxsi}[0]{{\it FOXSI}}

\newcommand{\app}[0]{$\sim$}
\newcommand{\per}[0]{$^{-1}$ }

\newcommand{\percub}[0]{$^{-3}$ }

\shorttitle{\nustar\, hard X-ray observation of a sub-A class solar flare}

\begin{document}

\title{\nustar\, hard X-ray observation of a sub-A class solar flare}

\author{Lindsay Glesener\altaffilmark{1}, S\"{a}m Krucker\altaffilmark{2,3}, Iain G. Hannah\altaffilmark{4}, Hugh Hudson\altaffilmark{2,4}, Brian W. Grefenstette\altaffilmark{5}, Stephen M. White\altaffilmark{6}, David M. Smith\altaffilmark{7}, Andrew J. Marsh\altaffilmark{7} }

\altaffiltext{1}{University of Minnesota, Minneapolis, USA}
\altaffiltext{2}{University of California at Berkeley, Berkeley, USA}
\altaffiltext{3}{University of Applied Sciences and Arts Northwestern Switzerland, Windisch, Switzerland}
\altaffiltext{4}{University of Glasgow, Glasgow, UK}
\altaffiltext{5}{California Institute of Technology, Pasadena, USA}
\altaffiltext{6}{Air Force Research Laboratory, Albuquerque, USA}
\altaffiltext{7}{University of California at Santa Cruz, Santa Cruz, USA}

\begin{abstract}
We report a \nustar\, observation of a solar microflare, SOL2015-09-01T04.  Although it was too faint to be observed by the \goes\, X-ray Sensor, we estimate the event to be an A0.1 class flare in brightness.  
This microflare, with only \app5 counts s\per detector\per observed by \rhessi, is fainter than any hard X-ray (HXR) flare in the existing literature.  The microflare occurred during a solar pointing by the highly sensitive \nustar\, astrophysical observatory, which used its direct focusing optics to produce detailed HXR microflare spectra and images. The microflare exhibits HXR properties commonly observed in larger flares, including a fast rise and more gradual decay, earlier peak time with higher energy, spatial dimensions similar to the \rhessi\, microflares, and a high-energy excess beyond an isothermal spectral component during the impulsive phase.  The microflare is small in emission measure, temperature, and energy, though not in physical size; observations are consistent with an origin via the interaction of at least two magnetic loops.  We estimate the increase in thermal energy at the time of the microflare to be 2.4$\times10^{27}$ ergs.  The observation suggests that flares do indeed scale down to extremely small energies and retain what we customarily think of as ``flarelike'' properties.  

\end{abstract}

\section{Introduction}

Solar flares are impulsive transformations of magnetic energy into heating, particle acceleration, and, occasionally, eruptions.  They are of interest for understanding the basic physics of the Sun because they represent a restructuring of the coronal magnetic field, are often accompanied by coronal mass ejections, accelerate a huge number of particles up to high energies \citep[e.g.][]{lin1976}, and may impulsively heat the corona \citep[e.g.][]{klimchuk2006}.  While many flare investigations concentrate on the largest events due to the rich multiwavelength observations and detailed phenomena that can be studied, there is also extensive investigative opportunity on the smaller side of the flare distribution, where flares are less dramatic but far more frequent.  In fact, if flares are responsible for coronal heating, it has been shown that this heating must be in weak flare-like events, not in the form of typical, larger flares \citep[e.g.][]{hudson1991}.

Hard X-rays (HXRs) are a useful tool in understanding energetics in flares of any size, because they are produced via bremsstrahlung by high-energy electrons that are either hot (millions to tens of millions of degrees) or nonthermal (accelerated by the flare).  For the smallest flares this can be especially key -- while the time necessary for ionization equilibrium means that line emission in extreme ultraviolet (EUV) or soft X-rays (SXR) might be suppressed for very short events \citep[e.g.][]{bradshaw2011}, HXR feedback is immediate.  

Thorough statistical and case studies were undertaken on HXR flares and larger microflares (also known as active region transient brightenings) using the Hard X-ray Telescope on the \textit{Yohkoh} spacecraft \citep{kosugi1991} in the 1990s and the \textit{Reuven Ramaty High Energy Solar Spectroscopic Imager (RHESSI)} \citep{lin2002} for solar cycles 23 and 24 (the current era).  These instruments employ indirect imaging and thus have been limited to flares of a certain brightness -- for \rhessi, temperatures $T\gtrsim$9 MK and emission measure $EM \gtrsim10^{45}$ cm$^{-3}$, with particularly good coverage of flares with $T\gtrsim$12 MK and $EM \gtrsim10^{46}$ cm\percub \citep{hannah2008}.  

\citet{nishio1997} examined 14 impulsive flares observed in microwaves (by Nobeyama) and SXRs/HXRs (by \textit{Yohkoh}) and found that 10 of the 14 displayed evidence for at least two loops.  The two loops were typically of very different lengths, of order $\lesssim$20 arcseconds and 30--80 arcseconds.  The HXRs and SXRs came predominantly from the more compact loop and its footpoints.  \citet{hanaoka1997} also identified a two-loop configuration in many radio/SXR flares, with many of the events showing a three-legged structure and large angles between the loops.  These observations have led to the modern concept that flares generally consist of multiple, independently excited, loop structures that may interact. 

\citet{battaglia2005} studied \rhessi\, flares from \goes\, class B to M classes, finding a spectral softening of the nonthermal electron distribution with smaller flare energy.  \citet{hannah2008} and \citet{christe2008} studied over 25,000 \rhessi\, microflares of \textit{GOES} class A and B, finding that they all arise in active regions and have properties similar to larger flares, including impulsive rises and slow decays and the presence of thermal and nonthermal spectral components.

At lower energies, active region microflares have been catalogued in soft X-rays by \citet{shimizu1995} using \textit{Yohkoh} data.  In the quiet Sun, small transient brightenings have been surveyed in the EUV using \textit{SOHO}/EIT \citep{benz2002} and \textit{TRACE} \citep{parnell2000, aschwanden2000}.  While these data sets are quite disparate in instrument, active region, and solar cycle timing, a rough power law is evident, with small events occurring far more frequently than large ones \citep[see, e.g. Figure 2 in ][]{hannah2011}.  It is yet unclear if the smallest events release enough energy to play a major role in heating the corona.

In recent years, new instruments have begun to demonstrate the dramatically increased sensitivity available via direct HXR focusing as opposed to \rhessi's indirect imaging method \citep{hurford2002}, with the first two flights of the \textit{Focusing Optics X-ray Solar Imager} \foxsi\, sounding rocket \citep{krucker2014, glesener2016} and occasional solar pointings by the \textit{Nuclear Spectroscopic Array} (\nustar) astrophysics spacecraft \citep{harrison2013, grefenstette2016}.  Focusing HXR instruments, with their larger effective areas and drastically reduced detector backgrounds, can measure flares of smaller temperatures, brightnesses, and total energies than those available to indirect imagers.  

We report here an observation of a small HXR microflare near the west limb on 2015 September 1 observed by \nustar.  Since the microflare produced only 5 \rhessi\ counts s\per detector\per (not enough to reconstruct an image), we believe it to be fainter than any HXR flare in the current literature.  

\begin{figure*}[htb]
\begin{center}
	\includegraphics[width=0.45\textwidth,angle=90]{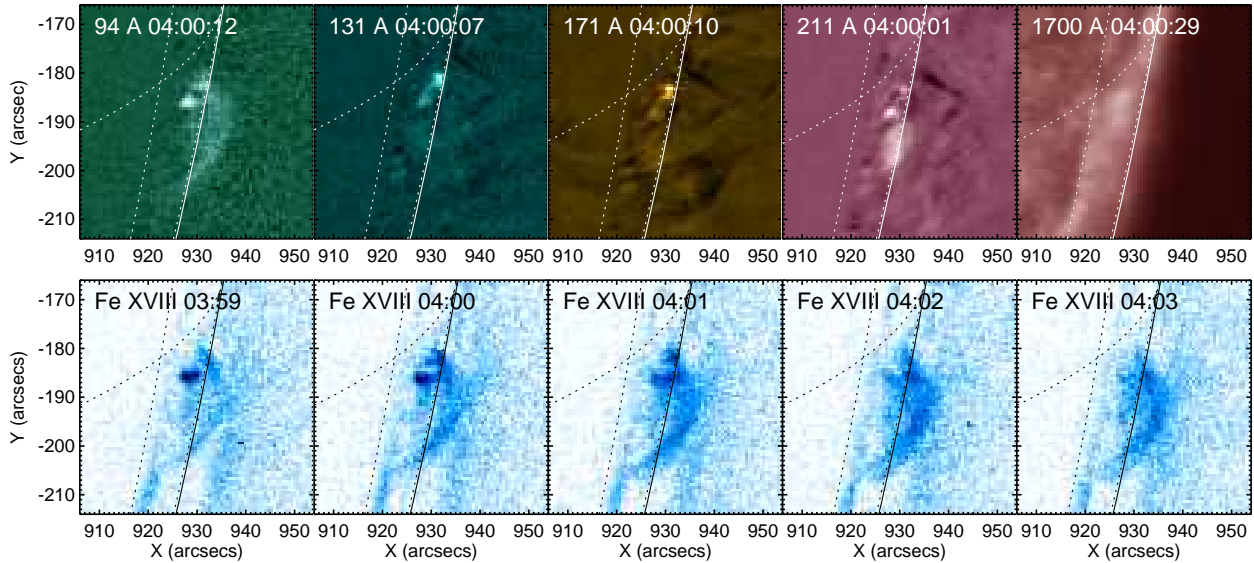}
\caption{\footnotesize AIA images of the 2015 Sep 01 microflare.  (Top row) View of the microflare in several different filters close to the peak time.  (Bottom row) estimates of the Fe XVIII contribution to AIA channels over time using a linear combination of 94\AA, 211\AA, and 171\AA\, emission, showing the evolution of hot plasma (\app4--10 MK).  }
\label{fig:aia}
\end{center}
\end{figure*}

\section{Observations}

\subsection{Overview of the event}
\label{sec:obs-overview}

\nustar\, is a NASA Small Explorer that uses directly focusing HXR optics to observe faint astrophysical sources \citep{harrison2013}.  While not explicitly designed for solar observing, \nustar's high-sensitivity telescopes can measure faint phenomena on the Sun during low-activity times, when best use is made of the instrument's limited throughput \citep{grefenstette2016}.  Solar pointings are coordinated as targets of opportunity and occur several times per year for a few orbits at a time.\footnote{Summary plots can be found at \url{http://ianan.github.io/nsigh_all/}.}  Prime conditions include (among other scenarios) a productive active region (AR) at the limb with an otherwise quiet disk.  This condition was met in early September 2015, and \nustar\, observed the Sun for 8 orbits spread out through the mornings of 2015 September 01 and 02.  The majority of the AR of interest, 12403, had occulted by that time, with only a small portion remaining on the visible disk.

Around 04:00 UT on 2015 September 01, a microflare occurred in the unocculted part of AR 12403.  The X-ray brightness of the microflare was below the sensitivity limit of the \textit{GOES} soft X-ray Sensor (XRS), which is typically used for flare brightness classification.  The microflare was observed by the Atmospheric Imaging Assembly (AIA) on the \textit{Solar Dynamics Observatory} \citep{lemen2012} and was independently identified in \nustar\, movies.  
Figure \ref{fig:aia} shows images from several of the AIA coronal bandpass filters (top row) and (bottom row) a sequence of images over time computed using a linear combination of three AIA filters (94\AA, 171\AA, and 211\AA) in order to estimate the Fe XVIII contribution (formation temperature $log(T)\approx6.9$), as in \citet{delzanna2013}.

\subsection{\rhessi\, HXRs}

The event was too faint to register in the \rhessi\, flare list, but manual inspection of the data identified a small count rate rise in the 4--9 keV range co-temporal with a \nustar\, peak in the same range.  Detector 1 shows the clearest detection of the nine \rhessi\, detectors since several other detectors had thresholds set too high to register the relatively low-energy flare.  Detector 1 emission reached a peak of 8.8 counts second$^{-1}$ over a background of 4.3 counts second$^{-1}$, for a total of \app120 photons, not enough to produce an image.

\begin{figure*}[htp]
\begin{center}
	\includegraphics[width=\linewidth]{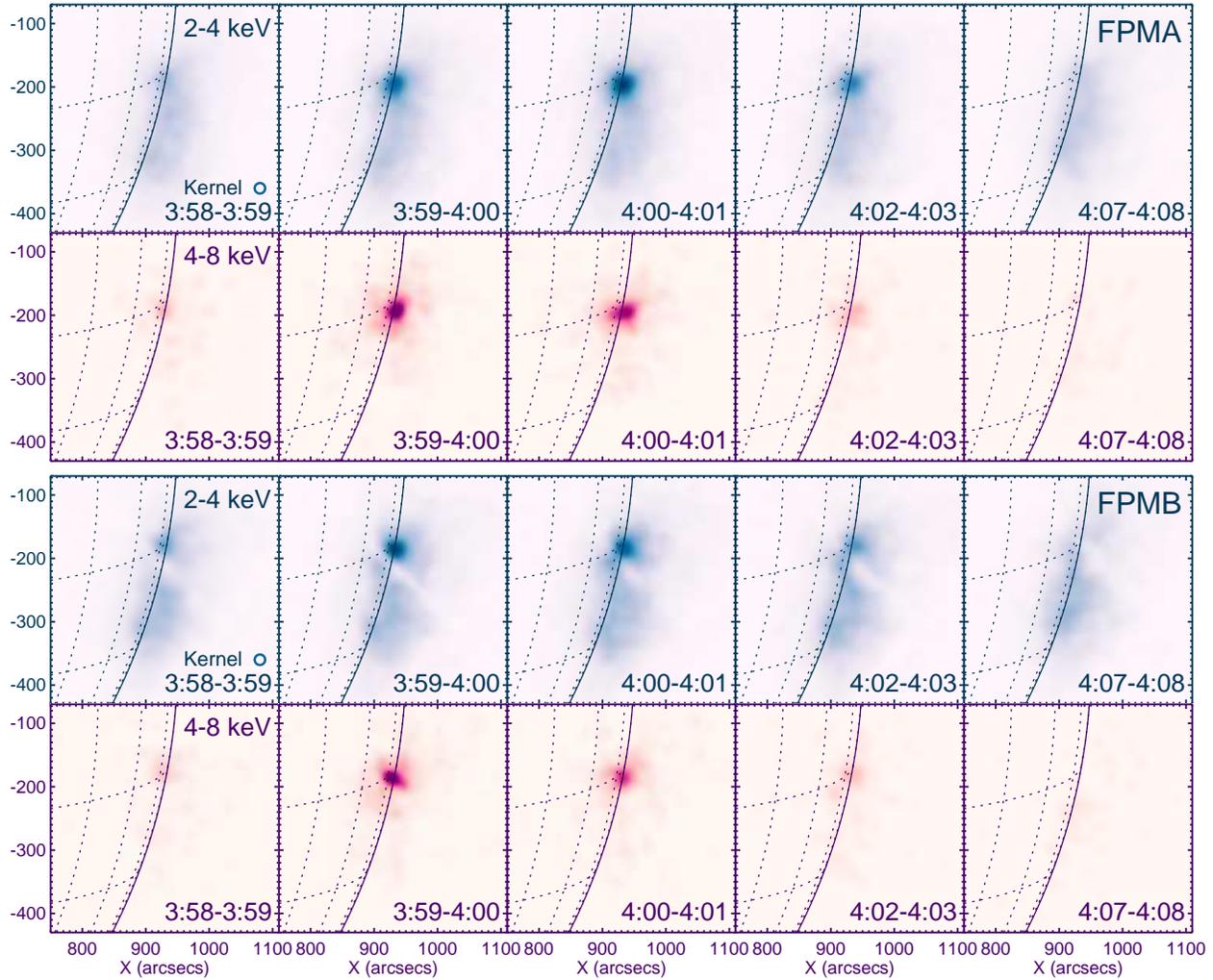}
\caption{\footnotesize \nustar\, images of the microflare in two energy bands.  The top two rows show images from FPMA at 2-4 keV and 4-8 keV, and the bottom two rows show the same for FPMB.  Images have been integrated for one minute, livetime-corrected, and smoothed over 17 arcseconds using a Gaussian kernel to reduce statistical noise.  The same intensity scale is used for all images in each row.  A diagonal gap across the source in the FPMB images is due to the space between detector quadrants.  By 4:07 UT (last column), the microflare has subsided and only quiescent emission is observed.  The 2-4 keV images show bright emission from the quiescent active region in addition to the microflare.}
\label{fig:nustar-images}
\end{center}
\end{figure*}

\subsection{\nustar\, HXR data and pointing corrections}

\nustar, with its high effective area (\app700 cm$^2$ at 5 keV) and minuscule detector background has higher sensitivity than ever before available at HXR energies \citep{harrison2013}.  However, as X-ray flux from active regions and flares is far larger than the instrument throughput (800 counts s$^{-1}$ maximum) we can generally only record a fraction of the incoming X-rays.  The livetime (defined as the fraction of time for which the detector is ready to acquire an event) when observing the Sun is typically limited to a small percentage \citep{grefenstette2016}.  For the 2015 September 01 microflare, the livetime for \nustar's Focal Plane Module A (FPMA; one of two \nustar\, detector arrays) was 1.57\% at a nonflaring time, dropping to 1.24\% at the peak.  The corresponding livetimes for Focal Plane Module B (FPMB) were 1.18\% and 1.01\%, respectively.  All data shown in this paper are corrected for livetime, which is measured on a one-second cadence, though error bars are derived from the raw counts.  The effect of this low livetime is that a nominally one-minute observation has an effective exposure time of less than one second.

\nustar's solar pointing is uncertain to $\lesssim$1.5 arcmin due to the forward-facing star camera being blinded by solar flux.  This pointing offset typically changes with orbital thermal changes.  (More sudden pointing changes due to changes in star camera combinations in use are not relevant for the time period of this microflare.)  In order to isolate the microflare region over time (even at nonflaring times), \nustar\, images integrated over 12 seconds were cross-correlated with AIA Fe XVIII images.  Cross-correlations were performed on a cadence of 12 seconds using an automated procedure.  A smooth curve was then interpolated through the points to determine the necessary \nustar\, pointing adjustment as a function of time.  This pointing correction relies on the assumption that the majority of the \nustar\, and AIA Fe XVIII emissions originate from the same location, a reasonable assumption given that \nustar\, is highly sensitive to the Fe XVIII contribution to the AIA 94\AA\, filter \citep[$log(T)\approx 6.5-7.2$; see ][]{lemen2012}.  The area used for cross-correlation includes the entire active region, although the microflare dominates at its peak time.  The pointing adjustments (of 28--69 arcsec) derived in this way have been applied to all images and regions selected for time profiles and spectra shown in this paper.

\subsection{\nustar\, HXR images, lightcurves, and spectra}
\label{sec:nustar}

\nustar\, images of the microflare over time are shown in Figure \ref{fig:nustar-images}.  FPMA and FPMB data are shown separately, and two energy bands (2--4 keV and 4--8 keV) are shown for each.  Images have been integrated for one minute, livetime corrected, and smoothed over 17 arcseconds using a Gaussian smoothing kernel.  A small space between detector quadrants produces a diagonal gap across the FPMB image.  By 4:07 UT (last column), the event has subsided and only quiescent emission is observed.  The 2-4 keV images show bright emission from the quiescent active region in addition to the microflare, while the 4--8 keV images show little emission except for the microflare itself.

\begin{figure*}[htb]
\begin{center}
	\includegraphics[width=0.455\linewidth,trim=0 -0.1cm 0 0cm,clip=true]{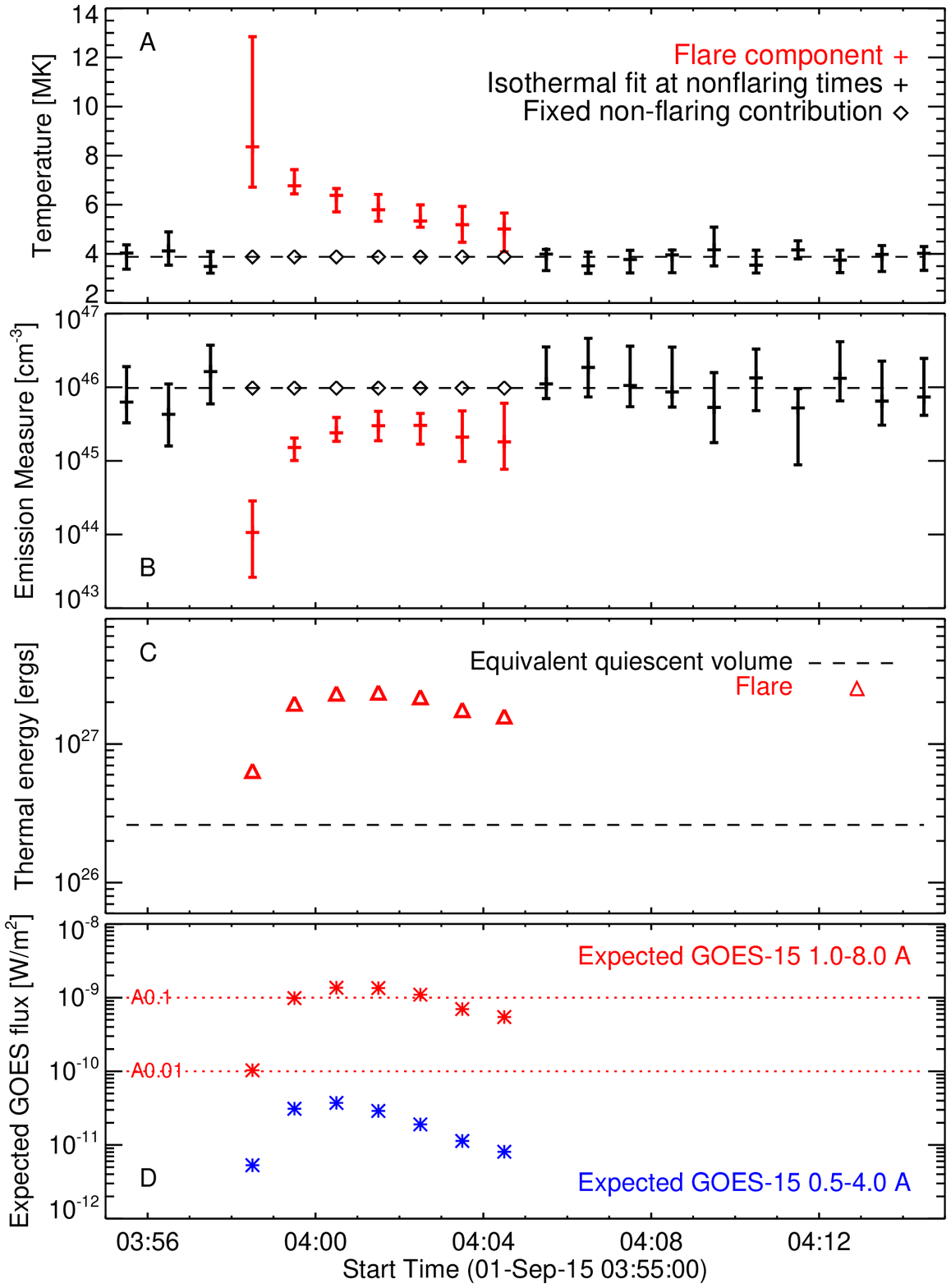}
	\includegraphics[width=0.45\linewidth, trim=0 0.8cm 2cm 0, clip=true]{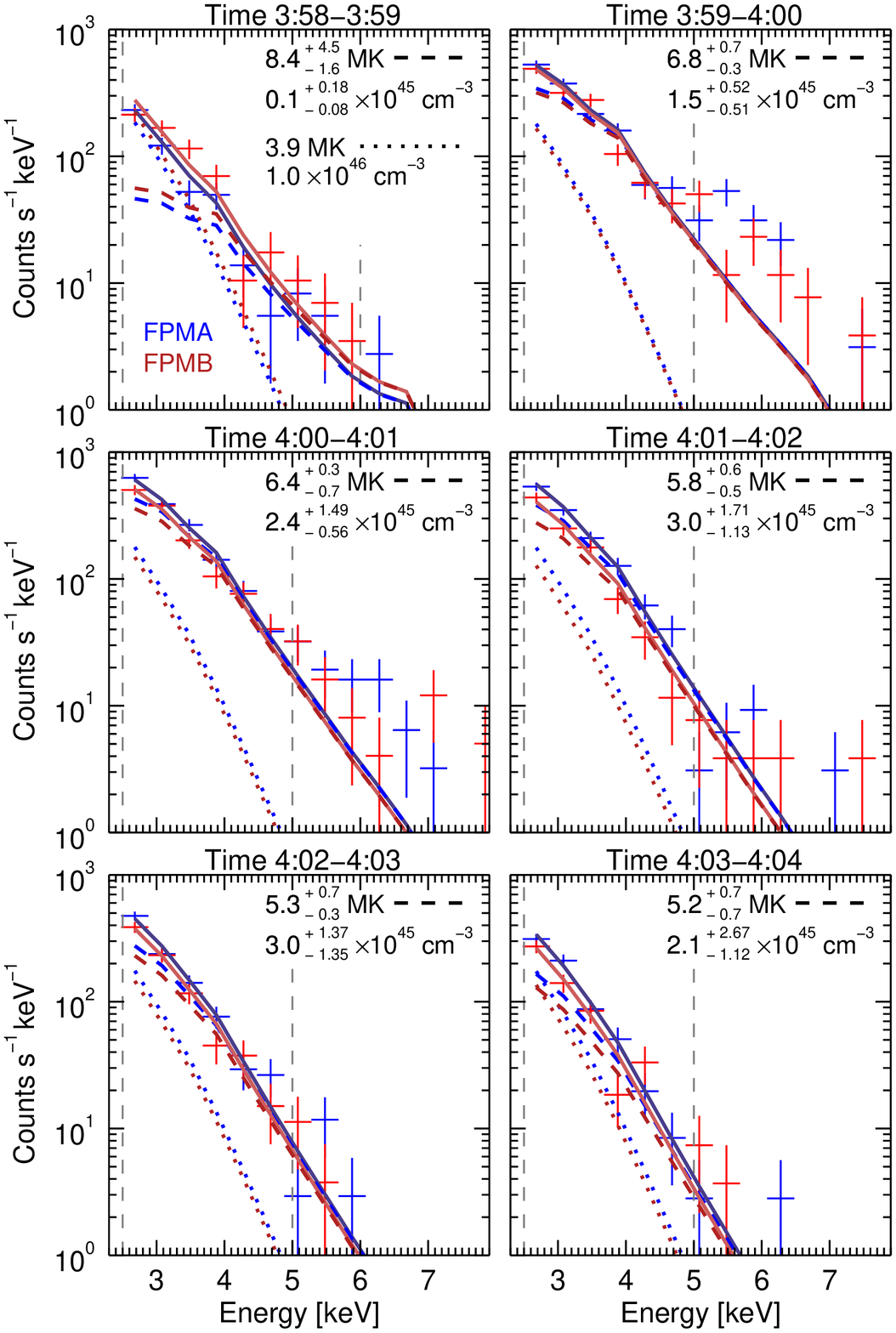}
\caption{\footnotesize (Lefthand panels) 
Evolution of \nustar\, spectral parameters throughout the microflare, using the spectral fits shown to the right.  Panels (a) and (b) show the fit temperatures and emission measures.  Panel (c) shows (red triangles) the thermal energy of the flaring volume compared with (black dashed line) the thermal energy of an equivalent quiescent volume based on fits at non-flaring times.  Panel (d) shows the \goes\, emission expected from the microflare, with A0.1 and A0.01 levels shown for comparison (red dotted lines).   
(Righthand panels) 
X-ray spectra fit simultaneously to FPMA and FPMB during the microflare.  Plots show FPMA and FPMB data points in blue and red, respectively, as well as (dotted line) a fixed thermal component at the quiescent level found in Section \ref{sec:nustar}, held identically for all flaring intervals, and (dashed line) a fit thermal component for the microflare.  The solid lines show the total fit models including all fit components.  Vertical dashed lines show the fit energy range.  In the second two intervals there is an excess in counts above the model above \app5 keV; this excess emission could be explained by a nonthermal component or by a small amount of much hotter plasma. 
}
\label{fig:thermal-evolution}
\end{center}
\end{figure*}

We fit the \nustar\, spectra using the XSPEC spectral fitting software \citep{arnaud1996}.  Fits were performed simultaneously to FPMA and FPMB data, and the pointing adjustments described in the previous section were applied in order to select a consistent region for fitting.  The region is a circle of radius 15 arcseconds centered at the microflare centroid location.  First, we fit an isothermal spectrum at 1-minute intervals throughout the time range shown in the lefthand side of Figure \ref{fig:thermal-evolution}.  From these values, we identified seven intervals that were obviously flaring.  We excluded the flaring intervals and computed the average temperature and emission measure for the nonflaring times, which we call the quiescent parameters.  Nonflaring fits are indicated with black markers and error bars in panels A and B of Figure \ref{fig:thermal-evolution}, and the average quiescent values are indicated with a dashed line.  Those quiescent, nonflaring parameters were then held as a fixed thermal component during the flaring times, while a second thermal component was fit to represent the microflare (see righthand side of Figure \ref{fig:thermal-evolution}).  The resulting thermal fits at flaring times are shown in red in panels A and B, while the fixed nonflaring component is shown with black diamonds (and black dashed lines).  

Next, we calculated the thermal energy $W_T$ of the microflare as $W_T=3 \sqrt{EM \, V}\, k_B T$, where $EM$ is the emission measure in cm$^{-3}$, $V$ is the microflare volume, $k_B$ is Boltzmann's constant, and $T$ is the temperature.  The volume was estimated from the AIA Fe XVIII images shown in Figure \ref{fig:aia} by considering the microflare loop to be a tube of roughly uniform radius.  We assumed the loop height to be perpendicular to the solar surface and corrected the loop length for the projection due to its near-limb position.  The resulting volume is \app3.2$\times 10^{26}$ cm$^3$.  
  We also calculated the thermal emission of an equivalent volume of quiescent plasma, i.e. using our quiescent background parameters and the same volume as the microflare.  In order to do this, we calculated a quiescent density from the (nonflaring) emission measure using the area of the region included in the \nustar\, spectroscopy (a circle of radius 15 arcseconds).  We approximated the line-of-sight extent from the longitudinal width of active region 12403 as gleaned from the NOAA history; the resulting dimension of 102 Mm is a slight overestimate.  The calculated quiescent density is 5$\times 10^8$ cm$^{-3}$.  
The thermal energy of the plasma volume throughout the microflare (red triangles) is shown in panel C of Figure \ref{fig:thermal-evolution}, along with a dashed black line that indicates the thermal energy of the equivalent volume of quiescent plasma at nonflaring times.  We find the microflare energy at its peak (2.4$\times 10^{27}$ ergs) to be greater than, but within an order of magnitude of, the quiescent energy.  A unity filling factor was assumed in all density estimates.  The quiescent energy we have calculated is a lower limit since we observe only at the highest temperatures; the likely presence of brighter but cooler plasma at quiet times would increase the quiescent thermal energy.

In order to compare with a common measure of flare brightness, we estimate the \goes\, XRS flux in the long and short wavelength bands that would be expected given our measured temperatures and emission measures; see panel D of Figure \ref{fig:thermal-evolution}.  For reference, the A0.1 and A0.01 levels (long wavelength channel) are indicated with red dotted lines.  The microflare peaks at an estimated \goes\, 1.0--8.0 \AA\, level of 1.4$\times$ 10$^{-9}$ W m$^{-2}$.    
  In actuality, the microflare is not observable by \goes\, in either channel due to the background contribution of the rest of the solar disk and due to sensitivity and/or sampling limitations.  We have also estimated the emission observable by the six AIA coronal filters, and find that the \nustar\, microflare peak brightness is consistent with the measured emission in the 94\AA\, filter, while the images in the other filters are dominated by their responses to plasma cooler than the 6 MK microflare.

\begin{figure*}[htb]
\begin{center}
	\includegraphics[width=0.48\linewidth]{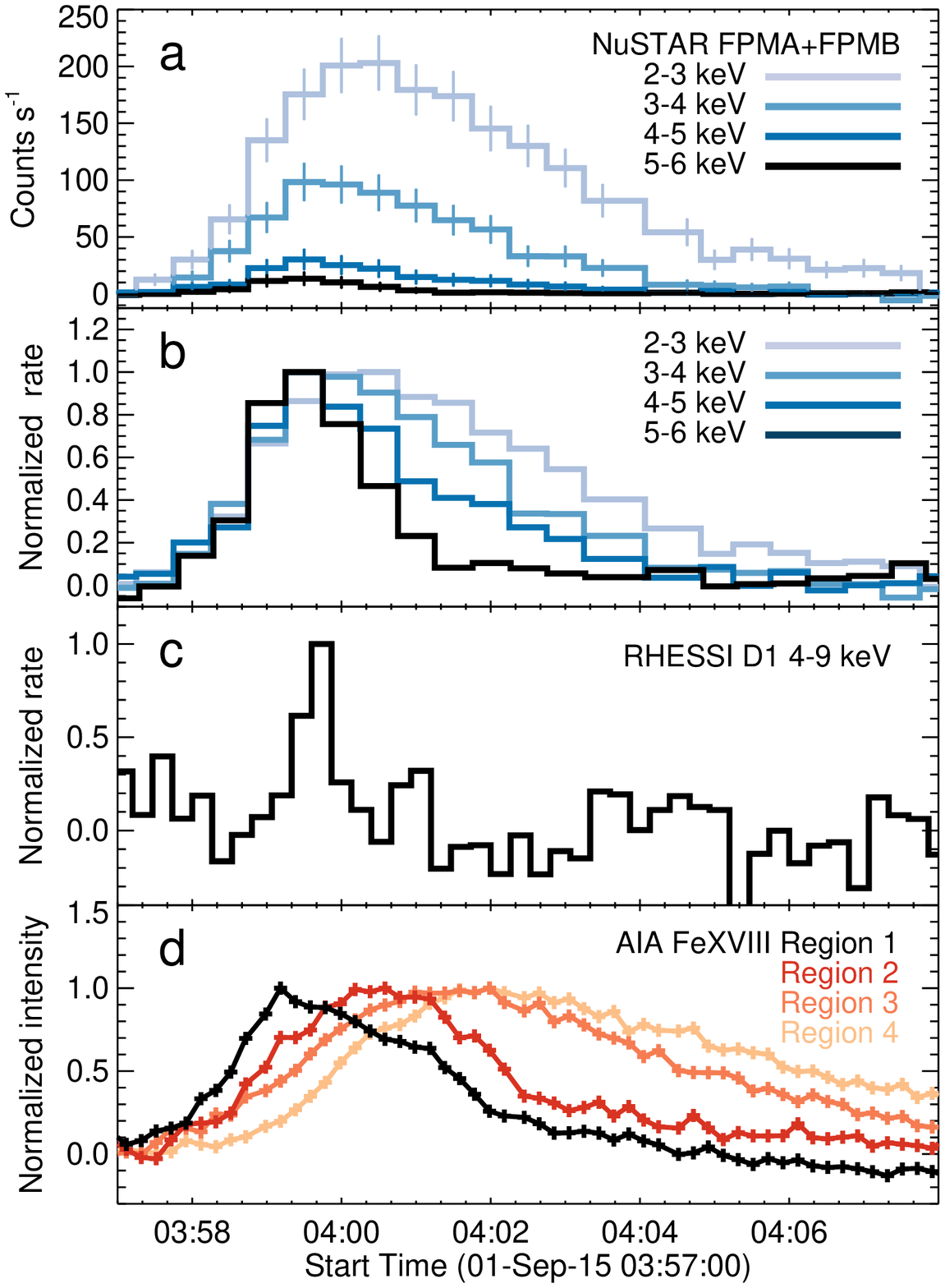}
	\includegraphics[width=0.35\linewidth]{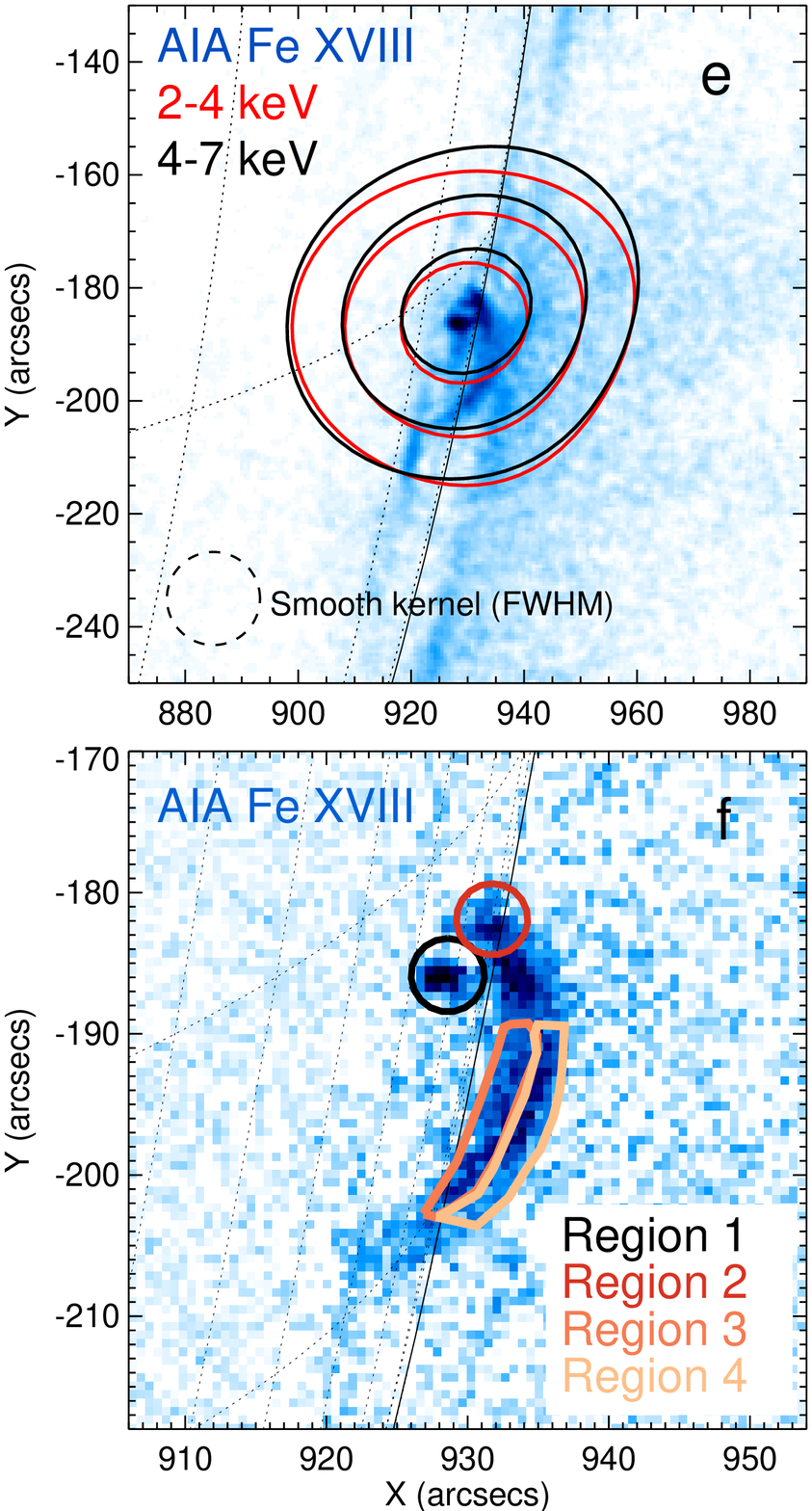}
\caption{\footnotesize (a) \nustar\, time profiles in 30-second bins for the entire active region in several energy bands.  Data from FPMA and FPMB have been livetime-corrected, background-subtracted, and added together.  (b) The same profiles, when normalized, show that the emission appears, on average, earlier with higher energy.  (c) \rhessi\, time profile for Detector 1 (D1) integrated over 4--9 keV. (d) Evolution of AIA Fe XVIII line emission for the four regions shown in Panel (f).  (e) \nustar\, contours, with preflare emission subtracted, coaligned to and overlaid onto a reference AIA Fe XVIII image (at 04:00 UT).  The \nustar\, images are integrated across the flare peak (0358--0402 UT) and have been smoothed over 17 arcseconds using a Gaussian kernel.  Contour levels are 50, 70, and 90\%.  (f) Selected AIA regions in a flare-integrated Fe XVIII image. }
\label{fig:lightcurve2}
\end{center}
\end{figure*}

\subsection{High-energy excess in the impulsive phase}

\nustar\, fits of binned count spectra for several intervals throughout the microflare are shown in the righthand side of Figure \ref{fig:thermal-evolution}.  Fits were performed simultaneously to FPMA and FPMB (blue and red data points, respectively) including a fixed thermal component (dotted lines) at the background level and a fitted thermal component for the microflare (dashed lines).  Isothermal flare components (plus the isothermal background component) fit the data well except during the impulsive phase, \app03:59--04:01 UT, where at high energies there is an evident excess in counts above the model.  This excess emission could be explained either by a nonthermal component due to flare-accelerated electrons with a rather flat power-law index of 3 or by a small amount of hotter plasma (temperature 13 MK, emission measure 3$\times$10$^{43}$ cm$^{-3}$).  However, both fits are poorly constrained given the low statistics above 5 keV.  

Since the pileup of photons arriving in quick succession could, in principle, produce a high-energy excess, we checked the pileup probability as indicated by the ``non-physical'' event grades; see Appendix C of \citet{grefenstette2016} for an explanation.  Since no events associated with the microflare were found to have non-physical grades, we conclude that pulse pileup does not affect our spectra.

Spectral fitting was also performed to \rhessi\, data using the OSPEX\footnote{\url{https://hesperia.gsfc.nasa.gov/ssw/packages/spex/doc/ospex_explanation.htm}} SSWIDL package.  Detectors 1, 3, and 5 detected enough photons to produce a spectrum.  A thermal fit to the spatially-integrated, detector-summed data result in a temperature of 13.1$\pm$4 MK and an emission measure of (1.7$\pm$3.2$)\times$10$^{44}$ cm$^{-3}$ during the time interval of 03:59--04:00 UT in the energy range 4--9 keV.  This is a somewhat hotter temperature than \nustar\, finds, but this is probably due to the high-energy excess present in this time interval.

\subsection{HXR and EUV evolution over time}

Figure \ref{fig:lightcurve2} examines the time behavior of the \nustar\, emission in various energy bands.  Panel (a) shows the evolution in 1-keV wide energy bins from 2 to 6 keV.  Data are summed from FPMA and FPMB and have been livetime-corrected and background subtracted.  These same lightcurves are shown, normalized to their respective maxima, in Panel (b).  The emissions show a slightly earlier peak time and faster decay with higher energy.  While data from FPMA and FPMB have been summed for better statistics, the FPMs individually show this trend.  Panel (c) shows 4-9 keV HXRs detected by \rhessi's Detector 1, though the emission is primarily $>5$ keV.  The small numbers of counts are statistically significant (5$\sigma$), and the time of the peak is roughly consistent with that of the higher \nustar\, energy bands.  
Panel (d) shows time profiles from four different regions of the AIA Fe XVIII images that are indicated in panel (f).  The most impulsive emission emanates from a compact source.  The northern section of the primary loop brightens next, and the lower and upper sections of the long loop show more gradual behavior.  Visual comparison of the \nustar\, and AIA lightcurves suggest that the HXR emission emanates from the compact sources in the northern region of the microflare, which are not resolved by \nustar.  See Panel (e) of Figure \ref{fig:lightcurve2}, though note that \nustar\, images have been coaligned to AIA.

\section{Discussion}

To summarize the observations, \nustar\, successfully observed an extremely weak, \app A0.1 class microflare that reached a peak thermal energy of 2.4$\times$10$^{27}$ ergs.  \nustar\, data show a clear trend of earlier peak time with higher HXR energies.  The temperature rises quickly and falls slowly, while the emission measure has a gradual rise and fall, suggesting impulsive energy release early in the event, followed by gradual filling and draining of the flare loop(s).  Spectral HXR fits show a high-energy excess in the impulsive phase.  All of these features are consistent with those observed in larger flares (e.g. those observed by \rhessi).

\begin{figure}[htbp]
\begin{center}
	\includegraphics[width=\linewidth, trim=0.3cm 0 1cm 0, clip=true]{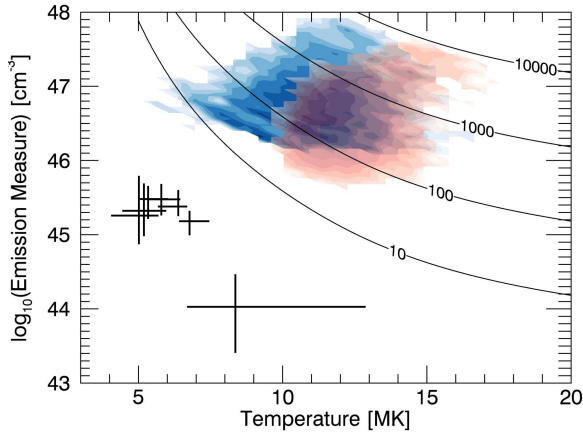}
\caption{\footnotesize \nustar\, spectral fit temperature and emission measure compared with \textit{GOES} (blue) and \rhessi\, (red) microflares studied by \citet{hannah2008}. Data points with error bars show the \nustar\, fit parameters for seven consecutive 1-minute intervals throughout the 2015 September 01 microflare.  Black contours give expected \rhessi\, counts s$^{-1}$ detector$^{-1}$. }
\label{fig:nustar_on_rhessi_em_t}
\end{center}
\end{figure}

Figure \ref{fig:nustar_on_rhessi_em_t} shows the \nustar\, temperature and emission measure at seven times throughout the microflare compared with \textit{GOES} (blue) and \rhessi\, (red) microflares studied by \citet{hannah2008} and \citet{christe2008}.  The \nustar\, microflare is cooler or fainter than all of the \rhessi\, microflares, and the temperature at the peak time (6.4 MK at 4:00--4:01 UT) is approximately half that of the typical \rhessi\, microflare (12.6 MK).  In further comparison, the SphinX instrument observed flares during its spatially-integrated observations in 2009, including two small flares with emission measures of $\sim$10$^{46}$cm$-3$ and temperatures of 4--5 MK \citep{engell2011}.  The \nustar\, microflare has an estimated loop length and volume of 70 Mm and 3.2$\times 10^{26}$ cm$^3$.  When compared with the loop lengths and volumes for the \rhessi\, microflares (Figure 4 in \citet{hannah2008}), we find that the small energy release takes place in a loop that is not unusually short or low-volume. 

Based on the \citet{shimizu1995} study of \textit{Yohkoh} SXR transients, the expected occurrence rate for microflares on the order of $10^{27}$ ergs is $10^{-52}$ s$^{-1}$ cm$^{-2}$ erg$^{-1}$, which works out to \app5.5 flares hour$^{-1}$ for the entire disk.  Dividing this rate between the four named active regions on the disk on 2015 September 01 yields an expected rate of \app1 flare hour$^{-1}$ active region$^{-1}$.  (Each \nustar\, orbit yields approximately one hour of solar observing time.)  No other AIA brightenings were visible by eye during the given orbit, although the microflare region does exhibit a precursor brightening just before the \nustar\, observation begins.  Therefore, our observation of a single microflare in this orbit is not unusual.

While the microflare is fainter than any previously reported HXR flare, it is clear that the event is \textit{not} a single energy release.  The AIA Fe XVIII emission displays complexity, with a small, compact source brightening first, followed by another, nearby, compact region.  Either or both of these compact sources could be small loops.  In the longer loop, brightening progresses from low to high altitudes.  This is commonly observed in larger flares and, in those cases, is usually interpreted as reconnecting of field at progressively higher altitudes within a flare arcade \citep[e.g.][]{gallagher2002}.

Where does high-energy emission lie within this dynamic picture?  The \nustar\, time profiles in Figure \ref{fig:lightcurve2} and the spectra in Figure \ref{fig:thermal-evolution} show an impulsive phase of the microflare from 03:59--04:00 UT.  It is here that the most significant high-energy \nustar\, emission is observed, along with a high-energy excess that cannot be accounted for by an isothermal model of the flaring emission.  This high-energy excess ($\gtrsim$5 keV in Figure \ref{fig:thermal-evolution}) could be provided by nonthermal electrons, as are commonly observed in the impulsive phase of larger flares; if present the nonthermal power law must be quite flat, with an index of $\sim$3.  Alternatively, the emission could be provided by a hot, faint, thermal component (temperature 13 MK, emission measure 3$\times$10$^{43}$ cm$^{-3}$).  
In either case, the \nustar\, emission is likely associated with energy release early in the microflare, and the compact AIA source discussed next responds immediately to this energy release.

\citet{melrose1997} considered a model of reconnection between current-carrying loops and found an energetically favorable configuration to be the reconnection of two current-carrying loops at large angles to one other, with closely-spaced footpoints of opposite polarity, so that the by-products are a small, compact loop and a longer, overarching one.  Our observations of this microflare are consistent with that geometry.  The compact source located at approximately [928,-185] arcseconds could be the (unresolved) short post-reconnection thermal loop.  This loop is hot but small, and could quickly fill with chromospheric plasma.  The longer loop may take longer to fill due to its length, accounting for the differing timescales in the lightcurves of the two loops.  Observationally, this microflare is consistent with the survey performed by \citet{nishio1997}, which found two-loop interactions in most of the 14 events they examined, as well as large asymmetries in the two loop lengths.

Studies of such small events address the relationship between nanoflares and flares. Since individual nanoflares are not currently observable, their contribution to coronal heating is often addressed by assuming that they are the low-energy end of a single power-law distribution of larger, already-observed flares \citep[see, e.g.,][]{hannah2011}. This extrapolation explicitly requires that the physics of nanoflares be the same as the physics of much larger flares. At a crude level, the mechanism for flares in the solar atmosphere has a source of free energy, believed to be in the form of magnetic fields, that accumulates slowly until it exceeds some threshold, at which time the energy is released impulsively. This is consistent with the original idea of \citet{parker1988} for nanoflares, in which current sheets arise due to footpoint motion and increase in strength until the coronal magnetic field change across the current sheet exceeds a threshold and triggers rapid magnetic reconnection. The question of a trigger is crucial: for many years magnetic reconnection theory focused on how to enable fast reconnection in order to match the energy release rates observed in large flares, but successfully doing so then raises the question of how to suppress reconnection while energy is being stored between flares (e.g., Cassak et al 2006).  An obvious difference between nanoflares and larger events is in the volume of magnetic field whose energy is released.  Nanoflares are assumed to occur on very small scales between almost parallel field lines, whereas the amount of energy released in a large flare requires a large volume for storage and could involve large field angles.  It is not clear that the same trigger mechanism should operate over such different regimes, and so it is not obvious that flares and nanoflares should form opposite ends of a single distribution.  It is therefore critical to push flare measurements to smaller and smaller scales in order to probe the transition between flare triggering regimes.

\section{Summary}

We have presented a \nustar\, microflare fainter than any in the previous HXR literature.  We estimate the microflare to be class $\sim$A0.1, with a peak thermal energy of 2.4$\times10^{27}$ ergs, similar to the quiescent energy of an equivalent plasma volume.  We observe several qualities common to larger HXR flares, such as early, impulsive energy release followed by a gradual thermal response, and even more gradual at higher altitudes.  HXRs peak earlier with higher energy and show a high-energy excess during the impulsive phase that is due to either nonthermal electrons or faint, hotter plasma.  We conclude that flares do indeed scale down to extremely small energies and retain what we customarily think of as ``flarelike'' properties.

We wish to emphasize that, while new to the literature, this microflare is \textit{not} unique.  Other \nustar\, microflares will be the subject of upcoming papers, including one by \citet{wright2017} that shows a detailed differential emission measure obtained via \textit{Hinode} XRT coordination.  We expect future \nustar\, events with higher livetime and correspondingly better statistics.  In addition, the \foxsi\, sounding rocket has observed three microflares in its first two flights \citep{glesener2016}, one of which is estimated to be an A0.5 class flare \citep{vievering2016}.  This set of observations suggests that as more sensitive instruments are developed, even smaller flares will be discovered, allowing more thorough understanding of flare energetics and triggering, and the relationship between flares, microflares, and nanoflares.  Thorough discovery of this new flare regime must wait until a solar-dedicated spaceborne mission with focusing HXR optics is realized.

\acknowledgments

Support for this work was provided by a NASA HSR grant NNX14AG07G, as well as an NSF Faculty Development Grant (AGS-1429512) to the University of Minnesota.  S.K. acknowledges funding from the Swiss National Science Foundation (200021-140308 and 200020-169046).  I.G.H. is supported by a Royal Society University Fellowship.  The authors are grateful to the \nustar\, Science and Operation teams for promoting and supporting solar observations, to Matej Kuhar and Kathy Reeves for insightful comments on paper drafts, and to the anonymous reviewer for excellent comments and questions that undoubtedly improved the paper.


\begin{thebibliography}

\bibitem[Arnaud(1996)]{arnaud1996} Arnaud, K.~A.\ 1996, Astronomical Data Analysis Software and Systems V, 101, 17 

\bibitem[Aschwanden et al.(2000)]{aschwanden2000} Aschwanden, M.~J., Tarbell, T.~D., Nightingale, R.~W., et al.\ 2000, \apj, 535, 1047 

\bibitem[Battaglia et al.(2005)]{battaglia2005} Battaglia, M., Grigis, P.~C., \& Benz, A.~O.\ 2005, \aap, 439, 737 

\bibitem[Benz \& Krucker(2002)]{benz2002} Benz, A.~O., \& Krucker, S.\ 2002, \apj, 568, 413 

\bibitem[Bradshaw \& Klimchuk(2011)]{bradshaw2011} Bradshaw, S.~J., \& Klimchuk, J.~A.\ 2011, \apjs, 194, 26 

\bibitem[Cassak et al.(2006)]{cassak2006} Cassak, P.~A., Drake, J.~F., \& Shay, M.~A.\ 2006, \apjl, 644, L145

\bibitem[Christe et al.(2008)]{christe2008} Christe, S., Hannah, I.~G., Krucker, S., McTiernan, J., \& Lin, R.~P.\ 2008, \apj, 677, 1385-1394 

\bibitem[Christe et al.(2011)]{christe2011} Christe, S., Krucker, S., \& Saint-Hilaire, P.\ 2011, \solphys, 270, 493 

\bibitem[Del Zanna(2013)]{delzanna2013} Del Zanna, G.\ 2013, \aap, 558, A73 

\bibitem[Engell et al.(2011)]{engell2011} Engell, A.~J., Siarkowski, M., Gryciuk, M., et al.\ 2011, \apj, 726, 12 

\bibitem[Feldman1992 et al.(1992)]{feldman1992} Feldman, U., Mandelbaum, P., Seely, J.~F., Doschek, G.~A., \& Gursky, H.\ 1992, \apjs, 81, 387 

\bibitem[Gallagher et al.(2002)]{gallagher2002} Gallagher, P.~T., Dennis, B.~R., Krucker, S., Schwartz, R.~A., \& Tolbert, A.~K.\ 2002, \solphys, 210, 341 

\bibitem[Glesener et al.(2016)]{glesener2016} Glesener, L., Krucker, S., Christe, S., et al.\ 2016, \procspie, 9905, 99050E 

\bibitem[Grefenstette et al.(2016)]{grefenstette2016} Grefenstette, B.~W., Glesener, L., Krucker, S., et al.\ 2016, \apj, 826, 20 

\bibitem[Hanaoka(1997)]{hanaoka1997} Hanaoka, Y.\ 1997, \solphys, 173, 319 

\bibitem[Hannah et al.(2008)]{hannah2008} Hannah, I.~G., Christe, S., Krucker, S., et al.\ 2008, \apj, 677, 704 

\bibitem[Hannah et al.(2016)]{hannah2016} Hannah, I.~G., Grefenstette, B.~W., Smith, D.~M., et al.\ 2016, \apjl, 820, L14 

\bibitem[Hannah et al.(2011)]{hannah2011} Hannah, I.~G., Hudson, H.~S., Battaglia, M., et al.\ 2011, \ssr, 159, 263 

\bibitem[Harrison et al.(2013)]{harrison2013} Harrison, F.~A., Craig, W.~W., Christensen, F.~E., et al.\ 2013, \apj, 770, 103 

\bibitem[Hudson(1991)]{hudson1991} Hudson, H.~S.\ 1991, \solphys, 133, 357 

\bibitem[Hurford et al.(2002)]{hurford2002} Hurford, G.~J., Schmahl, E.~J., Schwartz, R.~A., et al.\ 2002, \solphys, 210, 61 

\bibitem[Klimchuk(2006)]{klimchuk2006} Klimchuk, J.~A.\ 2006, \solphys, 234, 41 

\bibitem[Kosugi et al.(1991)]{kosugi1991} Kosugi, T., Makishima, K., Murakami, T., et al.\ 1991, \solphys, 136, 17 

\bibitem[Krucker et al.(2014)]{krucker2014} Krucker, S., Christe, S., Glesener, L., et al.\ 2014, \apjl, 793, L32 

\bibitem[Lemen et al.(2012)]{lemen2012} Lemen, J.~R., Title, A.~M., Akin, D.~J., et al.\ 2012, \solphys, 275, 17 

\bibitem[Lin et al.(2002)]{lin2002} Lin, R.~P., Dennis, B.~R., Hurford, G.~J., et al.\ 2002, \solphys, 210, 3

\bibitem[Lin \& Hudson(1976)]{lin1976} Lin, R.~P., \& Hudson, H.~S.\ 1976, \solphys, 50, 153 

\bibitem[Melrose(1997)]{melrose1997} Melrose, D.~B.\ 1997, \apj, 486, 521 

\bibitem[Nishio et al.(1997)]{nishio1997} Nishio, M., Yaji, K., Kosugi, T., Nakajima, H., \& Sakurai, T.\ 1997, \apj, 489, 976 

\bibitem[Parker(1988)]{parker1988} Parker, E.~N.\ 1988, \apj, 330, 474

\bibitem[Parnell \& Jupp(2000)]{parnell2000} Parnell, C.~E., \& Jupp, P.~E.\ 2000, \apj, 529, 554 

\bibitem[Shimizu(1995)]{shimizu1995} Shimizu, T.\ 1995, \pasj, 47, 251 

\bibitem[Vievering et al.(2016)]{vievering2016} Vievering, J.~T., Glesener, L., Krucker, S., et al.\ 2016, AGU Fall Meeting Abstract \#SH11D-05

\bibitem[Wright et al.(2017)]{wright2017} Wright, P.~J., Hannah, I.~G., Grefenstette, B.~W., et al.\ 2017; accepted. arXiv:1706.06108 


\end{thebibliography}
\end{document}